\documentclass[preprint,12pt]{elsarticle}

\usepackage{amsmath}
\usepackage{amssymb}

\usepackage{multirow}

\biboptions{comma,square,compress}

\journal{Placenta}

\begin{document}

\begin{frontmatter}



\title{Probability distributions for measures of placental shape and morphology}


\author[1]{J. S. Gill}
\author[1]{M. P. Woods}
\author[2]{C. M. Salafia}
\author[1]{D. D. Vvedensky}

\address[1]{The Blackett Laboratory, Imperial College London, London SW7 2AZ, United Kingdom}
\address[2]{Placental Analytics LLC, 93 Colonial Avenue, Larchmont, New York 10538}

\begin{abstract}
{\bf Goal.} Weight at delivery is a standard cumulative measure of placental growth. But weight is a crude summary of other placental characteristics, such as the size and shape of the chorionic plate and the location of the umbilical cord insertion.  Distributions of such measures across a cohort reveal information about the developmental history of the chorionic plate that is unavailable from an analysis based solely on the mean and standard deviation.

\noindent
{\bf Methods \& Materials.} Various measures were determined from digitized images of chorionic plates obtained from the Pregnancy, Infection, and Nutrition Study, a prospective cohort study of preterm birth in central North Carolina between 2002 and 2004.  The centroids (the geometric centers) and umbilical cord insertions were taken directly from the images. The chorionic plate outlines were obtained from an interpolation based on a Fourier series, while eccentricity (of the best-fit ellipse), skewness, and kurtosis were determined from a shape analysis using the method of moments. The distribution of each variable was compared against the normal, lognormal, and L\'evy distributions.

\noindent
{\bf Results.} We found only a single measure (eccentricity) with a normal distribution.  All other placental measures required lognormal or ``heavy-tailed'' distributions to account for moderate to extreme deviations from the mean, where relative likelihoods in the cohort far exceeded those of a normal distribution.

\noindent
{\bf Conclusions.} Normal and lognormal distributions result from the accumulated effects of a large number of independent additive (normal) or multiplicative (lognormal) events.  Thus, while most placentas appear to develop by a series of small, regular, and independent steps, the presence of heavy-tailed distributions suggests that many show shape features which are more consistent with a large number of correlated steps or fewer, but substantially larger, independent steps.
\end{abstract}


\end{frontmatter}



\section{Introduction}
\label{sec1}

The placenta is the interface across which all oxygen and nutrients are exchanged between mother and fetus.  Understanding the development and function of the human placenta is crucial to gaining insight into the environment of the developing fetus, whose health is thought to be an important influence on childhood and lifelong health \cite{barker95}.

The placenta is conventionally thought to develop uniformly outward from the site of the umbilical cord insertion, leading to an approximately circular shape.  However, while circular placentas are infrequently observed, recent work \cite{yampolsky08b} has suggested that the ``average'' placental shape within a cohort is, in fact, close to circular, though there remains some debate on this issue \cite{pathak10,nelson10}. The ability of the chorionic late to extend laterally uniformly outward from the cord insertion is due, in part, to the suitability of the maternal uteroplacental environment.  Any deficiencies in that environment can have adverse effects on placental and, by extension, fetal development.  Consequently, the analysis of the deviations of mature placental chorionic surface shapes from ``regularity'' (circular, or otherwise) can provide information about the uterine environment and possibly provide indicators about the health of the child.

The structure of the mature placenta is geometrically complex.  The umbilical cord is usually attached near the center of the placenta, but this is not always the case;~eccentric, marginal and velamentous cords inserted onto the extraplacental membranes are not rare \cite{benirschke06}.  From the point of the cord insertion onto the chorionic plate, the fetal chorionic vascular system branches and spreads laterally across the chorionic plate.  At later stages in the branching and extension of chorionic surface vessels, veins and arteries dive down into the placenta and continue branching to contribute to disk thickness.  The chorionic surface outline of the delivered placenta is a culmination of lateral placental vascular growth.

A mature placenta can take many different shapes, from near-circular to multi-lobed to star-shaped.  There is little or no explanation as to why such variations of placental shapes exist, apart from ``trophotropism'' \cite{ramos88,cunningham97,benirschke00}, an argument which says, in effect, that ``the placenta grows where it can and does not grow where it cannot''. There is no data about whether shape variations are associated with particular complications or subsequent health problems. One of the main goals of the present study is to understand the genesis, development, and evolution of mature placental chorionic surface shape from the distributions of various measures of placental shape.  The shape of a placental chorionic surface or, indeed, any two-dimensional object, can be characterized by area, perimeter, compactness (perimeter squared divided by area), eccentricity (of a bounding ellipse), elongation and rectangularity (of a bounding box), etc.  In addition, the chorionic plate outline can be analyzed in terms of its ``roughness'' and ``correlation'', both of which are standard measures used in the statistical analysis of rough surfaces \cite{barabasi95}.  The eccentricity and orientation of the best-fit ellipse, skewness, and kurtosis can be calculated from  the lower-order moments of the chorionic plate \cite{teague80,prokop92}, and the distance between the umbilical cord insertion and the centroid, which provides an indication about how the placenta developed with respect to the umbilical cord, is extracted directly from the images. The roughness and correlation function are based on a Fourier representation of each chorionic plate outline.

Although an ideal placental shape is expected to be regular, if only to minimize the cost of maintaining its vascular network, deviations from regularity can be quite pronounced, as noted above, and are not uncommon.    This indicates that the lateral growth of the placental chorionic surface is not typically a uniform process, but has an element of randomness in many, if not most pregnancies, that is, different regions of the placental chorionic surface may develop at different random rates.  The potentially important corollary to this is that the ``fetal programming'' hypothesis may be germane to the majority of births, since few placentas are round with perfectly central cords.  Various measures can and have been extracted from digitized images of placentas \cite{salafia05} and plotted as distributions, but an analysis of their distributions has yet to reported. Our fundamental premise is that the form of these distributions can provide information about the statistical properties of these measures that encode the underlying developmental properties that led to these distributions.  Attaining a better understanding of the timing of development of placental chorionic surface shape features, which may reflect early perturbations of placental vascular growth \cite{yampolsky08b} may clarify how risk of the wide range of diseases that have been associated with gestational pathology develops,  or when in  subsequent pregnancies of that mother surveillance might be expected to be useful in identification of recurrence, since there is a low but finite risk of recurrence after preelampsia \cite{mcdonald09}, preterm birth \cite{iams09}, fetal growth restriction \cite{kinzler07}, stillbirth \cite{reddy07}, or even miscarriage \cite{flint96,regan00}.

Placental growth has been shown to be empirically modeled by growth of a fractal by diffusion-limited aggregation \cite{yampolsky08a}. From this basic observation, we can consider the notion of a random walk \cite{feller68}, where a ``walker'' takes small sequential steps to the left or right, each chosen randomly with equal probability. As the number of steps increases, the distribution of possible distances from the walker's initial position approaches a normal distribution.  An alternative version of a random walk is based on independent random \emph{relative} increments which, as the number of steps increases, leads to a lognormal distribution \cite{limpert01}.  Finally, a random walk with step sizes that decay as a power law for large step lengths is known as a ``L\'evy flight''. The likelihood of a large step is much greater than for a random walk, which has the effect of enhancing the rate of displacement compared to a random walk, and the resulting displacements follow the L\'evy distribution \cite{tsallis97}.

\section{Methods \& Materials}
\label{sec2}

\subsection{The Placental Cohort}
\label{sec2.1}

The data set for our analysis is obtained from the digital images of placentas collected from the Pregnancy, Infection, and Nutrition Study, a cohort study of women recruited at mid-pregnancy from an academic health center in central North Carolina. The study population and recruitment techniques are described in detail elsewhere \cite{kaufman03}. Beginning in March 2002, all women recruited into this study were requested to consent to a detailed placental examination. As of October 1, 2004, 1159 women (94.6\%) consented to such examination and 1014 (87.4\%) had placentas collected and photographed for image analysis. Of these, 1008 (99\%) were suitable for analysis.

Placental gross examinations, histology reviews, and image analyses were performed at EarlyPath Clinical and Research Diagnostics, a New York State-licensed histopathology facility under the direct supervision of Dr.~Carolyn Salafia. The institutional review board from the University of North Carolina at Chapel Hill approved this protocol. The fetal surface of each placenta was wiped dry and placed on a clean surface, after which the extraplacental membranes and umbilical cord were trimmed from the placenta.

The fetal surface was photographed with the laboratory identification number together with 3~cm of a plastic ruler in the field of view using a standard high-resolution digital camera with a minimum image size of 2.3 megapixels. A trained observer captured the $(x,y)$ coordinates that marked the site of the umbilical cord insertion and a series of such points along the perimeter of the fetal surface. The perimeter coordinates were captured at intervals no greater than 1~cm, with additional coordinates if it appeared essential to accurately capturing the shape of the fetal surface.

\subsection{Measures of Chorionic Plate Shape and Morphology}
\label{sec2.2}

A Fourier series (\ref{appendixA}) is used to interpolate between the discrete points captured along the perimeter of the chorionic plate (Sec.~\ref{sec2.1}), resulting in a smooth outline. A Fourier series is a sum of trigonometric functions (sines and cosines) whose coefficients measure the deviation of the outline from circularity.  If only small deviations are present, then the first few terms in this series are sufficient.  But more terms are required to capture an outline that has rapidly-varying features, such as lobes and protrusions. This interpolation is used to calculate moments of the region surrounded by the outline (\ref{appendixB}).  With increasing order, these moments provide successively more detailed information about the shape and morphology of the chorionic plate.  Both the Fourier coefficients of the outline and the moments of the region surrounded by the outline are used to calculate measures of the shape and morphology of the chorionic plate.  Table~\ref{table1} summarizes the measures and their formulas.

\begin{table}[t!]
\caption{Measures of the chorionic plate that are calculated in this paper.  Against the name of each measure is its symbol, definition, and a formula expressed in terms of moments $\mu_{ij}$ of the region bounded by the chorionic plate outline (\ref{appendixB}) or the Fourier coefficients $a_n$ and $b_n$ of the outline (\ref{appendixA}). The fundamental mathematical definitions of these measures, from which the formulas in this table are derived, are given in Appendices A and B.}
\label{table1}
\bigskip
\centering\footnotesize
\newcommand\T{\rule{0pt}{2.6ex}}
\newcommand\B{\rule[-1.2ex]{0pt}{0pt}}
\newcommand\TT{\rule{0pt}{4ex}}
\newcommand\BB{\rule[-3ex]{0pt}{0pt}}
\begin{tabular}{|c|c|c|c|}
\hline
Name & Symbol & Definition & Formula\T\B\\ 
\hline\hline
\multirow{2}{*}{Area} & \multirow{2}{*}{$A$} & {\footnotesize Area within chorionic} & \multirow{2}{*}{$\mu_{00}$}\TT\\ 
& & {\footnotesize plate outline} & \BB\\
\hline
\multirow{3}{*}{Centroid} & \multirow{3}{*}{$(x_c,y_c)$} & {\footnotesize Geometric center of} & \multirow{3}{*}{$\displaystyle{\biggl({\mu_{10}\over\mu_{00}},{\mu_{01}\over\mu_{00}}\biggr)}$}\TT\\
& & {\footnotesize area within chorionic} & \\
& & {\footnotesize plate outline} & \BB\\
\hline
\multirow{2}{*}{Eccentricity} & \multirow{2}{*}{$e$} & {\footnotesize Eccentricity of} & \multirow{2}{*}{$\displaystyle{\sqrt{1-{b^2\over a^2}}}$} \TT\\
& & {\footnotesize bounding ellipse } & \BB\\
\hline
\multirow{3}{*}{Skewness} & \multirow{3}{*}{$(S_x,S_y)$} & {\footnotesize Asymmetry of image} & \multirow{3}{*}{$\displaystyle{\left({\mu_{30}\over\mu_{20}^{\,\,3/2}},{\mu_{03}\over\mu_{02}^{\,\,3/2}}\right)}$}\TT\\
& & {\footnotesize projections onto} $x$- & \\
& & {\footnotesize and $y$-axes} &\BB\\
\hline
\multirow{4}{*}{Kurtosis} & \multirow{4}{*}{$(K_x,K_y)$} & {\footnotesize Peakedness relative to }& \multirow{4}{*}{$\displaystyle{\left({\mu_{40}\over\mu_{20}^3}-3,{\mu_{04}\over\mu_{02}^3}-3\right)}$}\TT\\
& & {\footnotesize normal distribution of} & \\
& & {\footnotesize image projections onto} & \\
& & {\footnotesize $x$- and $y$-axes} & \BB\\
\hline
\multirow{3}{*}{Roughness} & \multirow{3}{*}{$W$} & {\footnotesize Standard deviation of} & \multirow{3}{*}{$\displaystyle{\Biggl[{1\over2}\sum_{n=1}^N\bigl(a_n^2+b_n^2\bigr)\Biggr]^{1/2}}$}\TT\\
& & {\footnotesize chorionic outline from} & \\
& & {\footnotesize the average radius} & \BB\\
\hline
\multirow{2}{*}{Correlation} & \multirow{3}{*}{$C(s)$} & {\footnotesize Standard deviation of} & \multirow{3}{*}{$\displaystyle{\Biggl[2\sum_{n=1}^N\bigl(a_n^2+b_n^2\bigr)\sin^2\biggl({\pi s n\over L}\biggr)\Biggr]^{1/2}}$} \TT\\
\multirow{2}{*}{Function} & & {\footnotesize chorionic outline as a} & \\
& & {\footnotesize function of separation} & \BB\\
\hline
\end{tabular}
\end{table}

The area bounded by the chorionic outline provides a cumulative measure of the development of the placenta at delivery. No information is provided about the shape or morphology of the chorionic plate -- this is contained in higher moments.  The skewness measures the asymmetry with respect to the mean of projections of the image onto the $x$- and $y$-axes, viewed as distributions.  Similarly, the kurtosis measures the peakedness or flatness of these projections relative to that of a normal distribution, whose kurtosis has the value 3.  Thus, a positive (resp., negative) kurtosis means that the distribution is more (resp., less) peaked than a normal distribution.  Each chorionic plate has also been represented by a ellipse, whose eccentricity and orientation are determined by the zeroth and first moments.

The chorionic plate outline provides complementary information to the moment analysis. The two measures we use are the roughness and the correlation function.  The roughness, defined in (\ref{eqA6}), is an average over the chorionic plate outline of root-mean-squared deviations from an average radius.  Thus, roughness measures the ``width'' of the deviations of the outline from a circle.  A small roughness indicates a narrow width, which corresponds to an approximately circular outline, while a large width results from larger deviations from circularity, such as those of lobed or star-shaped outlines.

A related measure of the irregularity of the outline is the correlation function, defined in (\ref{eqA8}) as the standard deviation of differences between all pairs of radii on the chorionic plate outline at a fixed separation.  Whereas the roughness measures the deviation from circularity by summing individual points along the outline, yielding a {\it number}, the correlation function involves differences between radii at a fixed separations along the outline, which is expressed as a {\it function} of this separation.  Thus, the correlation function is measure of roughness that is spatially resolved along the chorionic plate outline.  In this paper, however, we will not discuss the spatial dependence of the correlation function, but focus on its average properties.

\subsection{Probability Distributions}
\label{sec2.3}

When calculated for all of the placentas in our cohort, the measures compiled in Table~\ref{table1} yield  ranges of values that can be represented as distributions, that is, the relative frequencies of occurrences of the outcomes of the measures.  These distributions embody information about the developmental characteristics of placentas, which can be identified by comparing them with distributions that are associated with particular types of processes.  The distribution functions that we use in this paper are summarized below, with details provided in \ref{appendixC}.

The most common probability distribution is the Gaussian, or {\bf normal}, distribution.  The probability density of this distribution is completely characterized by its mean $\mu$ and standard deviation $\sigma$. Normal distributions are so common because of the central limit theorem, which states that such distributions are the cumulative result of a large number of {\it additive} random events \cite{feller68}.  A related distribution is the {\bf lognormal}, which is the probability of a variable whose {\it logarithm} is normally distributed \cite{limpert01}. The lognormal is a skewed distribution, which occurs when averages are low, variances comparatively large, and values of the quantity being measured cannot be negative. This distribution is the cumulative result of a large number of {\it multiplicative} random events.

A distribution that is qualitatively different from the normal and lognormal distributions is the symmetric {\bf L\'evy} distribution \cite{tsallis97}.  The main distinguishing characteristic of L\'evy distributions
is that the probability of extreme variations decays like a {\it power} of that variation, as indicated in (\ref{eqC7}).  Hence, the occurrence of such variations is far more likely than for a normal distribution, which decays {\it exponentially}. For this reason, L\'evy distributions are called ``heavy tailed.''  L\'evy distributions arise from additive random events which may involve quite large changes.  In contrast, the events that result in the normal and lognormal distributions are comparatively small.

\begin{figure}[t!]
\centering
\includegraphics[width=8cm]{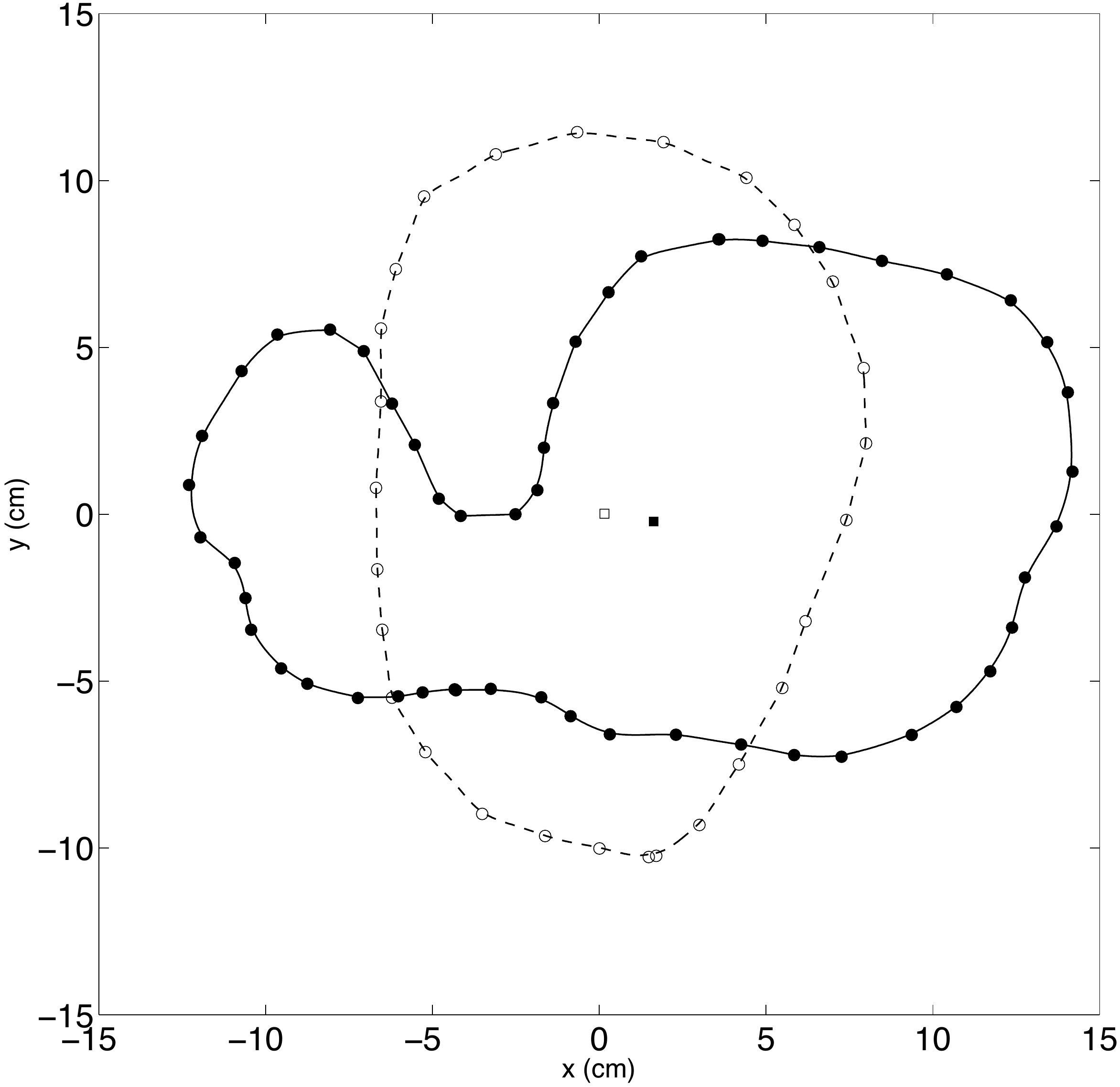}
\caption{Two examples of interpolations of chorionic plate outlines that have been determined by the method described in \ref{appendixA}.  The origin of the data points for each outline has been shifted to its centroid. The open circles mark the original data points and the broken curve shows the interpolation for an outline with a single-valued radius function.  The corresponding umbilical cord insertion is indicated by the interior open circle.  The closed circles mark the original data points and the solid curve shows the interpolation for an outline with a multi-valued radius function.  The corresponding umbilical cord insertion is indicated by the interior closed circle.}
\label{fig1}
\end{figure}

\section{Results}

\subsection{Interpolation of Chorionic Plate Outlines}
\label{sec3.1}

Figure~\ref{fig1} shows typical fits to the data points of chorionic plate outlines obtained with the method described in \ref{appendixA}.  Two types of outlines are shown:~one with a single-valued and one with a multi-valued radius.  A single-valued radius means that a line emanating from the centroid intersects every point on the perimeter only once, while a multi-valued radius function may intersect the perimeter more than once.  In the latter case, the perimeter folds back on itself, and the corresponding chorionic plate has lobes or some other irregular shape.  Note the irregular spacing of the points along the perimeter, as described in Sec.~\ref{sec2.1}.   The outline with the single-valued radius has a regular shape, so relatively few data points are needed.  However, the outline with the multi-valued radius has intervals where more points are needed to describe regions of greater curvature, which can occur for a small protrusion or, as in this case, a large morphological entity such as a lobe.  This is reflected in the number of terms that must be included in the Fourier series to produce an accurate interpolation.  The series for the outline with the single-valued radius required fewer terms than that for the outline with the multi-valued radius because regions of larger curvature mean that more rapidly varying trigonometric functions must be included in the interpolation.

\begin{figure}[t!]
\centering
\includegraphics[width=0.9\textwidth]{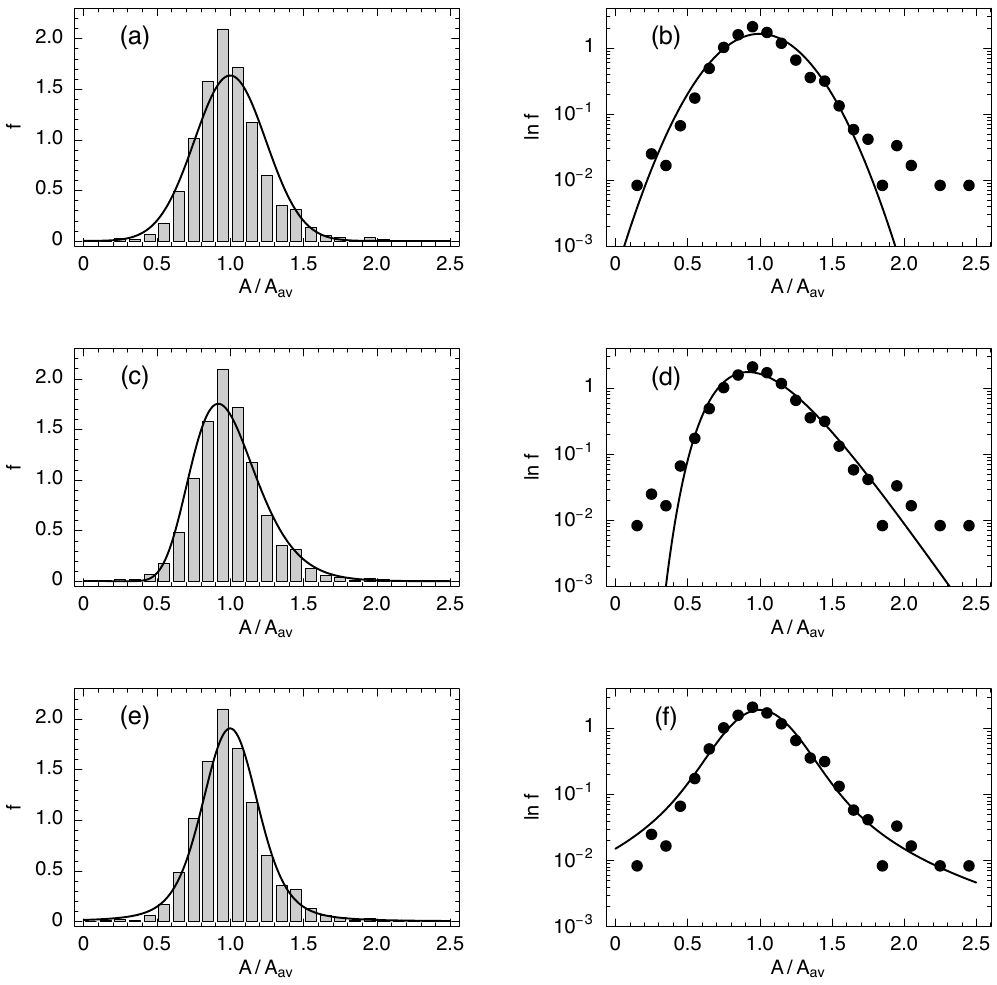}
\caption{Histogram of chorionic plate areas shown as the relative frequencies of bins of normalized areas.  These are compared with (a,b) the normal distribution, (c,d) the lognormal distribution, and (e,f) and an optimized L\'evy distribution, each of which is shown by a solid curve. The histogram and distributions are plotted on a logarithmic scale for the relative frequencies, indicated as points, in (b,d,f).}
\label{fig2}
\end{figure}

\subsection{Chorionic Plate Area}
\label{sec3.2}

The distribution of areas $A$ bounded by the outlines of the chorionic plates are shown as a histogram in Fig.~\ref{fig2}.  These histograms were constructed by first defining normalized areas as the original areas $A$ divided by the average area $A_{\rm av}$ of the cohort. These data points are grouped into contiguous ``bins'' of width 0.1, a choice dictated by the balance between the inherent statistical fluctuations in such a limited sample against the smoothness of the resulting relative frequency profile.  Choosing a width of 0.05 produced a somewhat noisier distribution but did not substantially alter any of the fits.  The relative frequencies $f$ are obtained by dividing the fraction of the total number of data points within each bin by the bin width, so the shaded area in the histogram in Fig.~\ref{fig2} is equal to 1.  This way of plotting histograms, which eliminates the units of the quantities being plotted, allows distributions of different measures to be compared directly, as well as providing the conceptual convenience of having the mean at 1.

Superimposed on the area histogram are the normal (a) and lognormal (b) distributions with mean and standard deviation determined from the data, in the latter case using (\ref{eqC3}) and (\ref{eqC4}), and an optimized fit to a L\'evy distribution (c), which yielded the parameters $\alpha=1.62\pm0.02$ and $\gamma=0.046\pm0.04$ in (\ref{eqC5}).  This fit was obtained by the least squares method, wherein a L\'evy distribution was calculated at the center of each bin, and the sum of the squares of the differences between these values and those of the bins was minimized by varying $\alpha$ and $\gamma$.  Figures~\ref{fig2}(b,d,f) show the same distributions plotted on a logarithmic scale for the frequencies (but maintaining the same linear scale for the areas).  Such plots are used to accentuate the extreme variations of data (the ``tails'' of the relative frequencies) to assess how various distributions account for this regime. Note that, according to (\ref{eqC7}), the fit in Fig.~\ref{fig2}(c), yields a probability for large deviations from the mean decreases as $(A/A_{\rm av})^{-2.62}$.

\subsection{Perimeter Roughness and Correlation Function}
\label{sec3.3}

\begin{figure}[t!]
\centering
\includegraphics[width=0.9\textwidth]{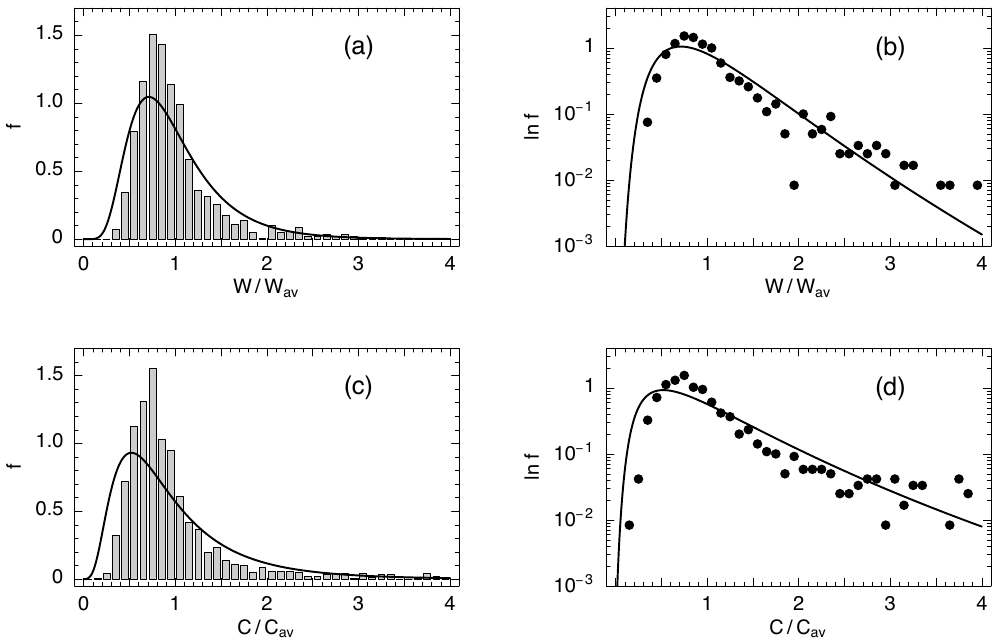}
\caption{Histograms of the roughness (a,b) and the integrated correlation function in (\ref{eqA10}) (c,d). Linear plots are shown in (a) and (c) and the corresponding plots with a logarithmic frequency scale in (b) and (d) with the frequencies associated with each bin indicated by points. Each of the histograms is compared with a lognormal distribution, which is indicated by the solid curve.}
\label{fig3}
\end{figure}

Measures of the chorionic plate outline that provide information about shape are the roughness of the perimeter(\ref{eqA6}) and the integrated correlation function (\ref{eqA10}).  Figure~\ref{fig3}(a,c) shows the histograms of these quantities, with each normalized as in Fig.~\ref{fig2}, i.e.~each measure is divided by its average over the cohort and the frequencies are defined such that the sum of the shaded regions is equal to 1.  The bin width for each histogram was again taken as 0.1.  Plotted with each histogram is a lognormal distribution whose mean and standard deviation were determined from the data by using (\ref{eqC3}) and (\ref{eqC4}).  Figure~\ref{fig3}(b,d) shows the histograms and corresponding lognormal distributions plotted on a logarithmic scale for the frequencies.  The tails of these distributions extend to much larger values than the area distributions in Fig.~\ref{fig2}, so the semi-logarithmic plots provide correspondingly more information about the distribution. These histograms are significantly skewed, so only the lognormal distribution is appropriate, as both the normal and L\'evy distributions are symmetric.

\subsection{Distance between the Centroid and the Umbilical Cord Insertion}
\label{sec3.4}

\begin{figure}[t!]
\centering
\includegraphics[width=0.9\textwidth]{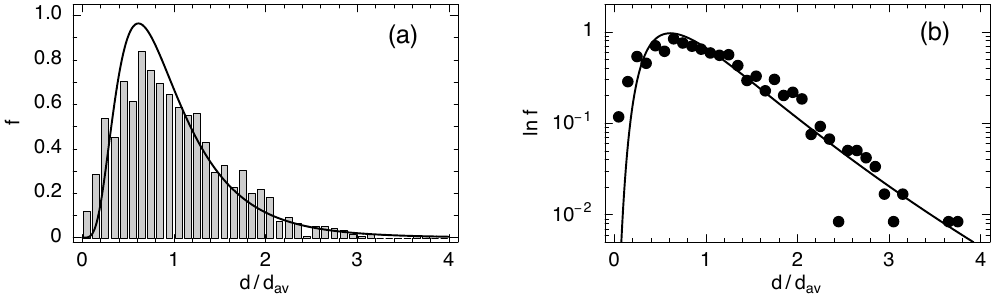}
\caption{(a) Histogram of the distances between the centroid and the umbilical cord insertion compared with the lognormal distribution.  (b) Semi-logarithmic plot of the histogram and distributions in (a), with the bin frequencies represented by points.}
\label{fig4}
\end{figure}

As the placenta grows outwards from a central point, the position of the umbilical cord insertion relative to the centroid of the placenta, which is a measure of the centrality of this point, provides information about the isotropy of placental development.  If the umbilical cord insertion is close to the centroid, then the placenta has, on average, grown outwards more symmetrically than if the cord insertion is displaced appreciably from the centroid. This does not imply that the chorionic plate is circular in this case, just that lateral growth was not skewed in any direction.  The histogram of the distances between the centroid and the umbilical cord insertion is show in Fig.~\ref{fig4}, plotted with the distances divided by their average, with frequencies that sum to 1.  The data have been grouped into bins of width 0.1.  Superimposed on the histograms are the lognormal distribution whose mean and standard deviation are determined from the data by using (\ref{eqC3}) and (\ref{eqC4}).  Note that, in common with the histograms in Fig.~\ref{fig3}, the histogram of the distances is highly skewed, with a long tail, so only the lognormal distribution is appropriate.

\begin{figure}[p!]
\centering
\includegraphics[width=13cm]{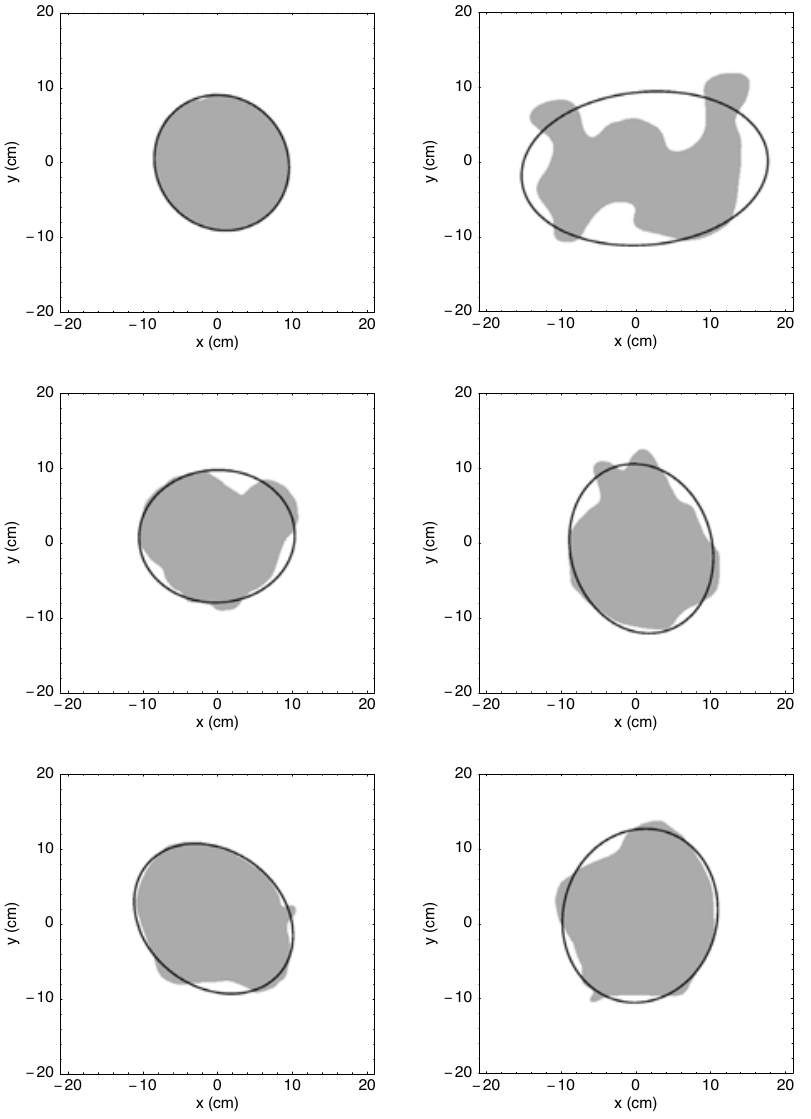}
\caption{Chorionic plates (shown shaded) and best-fit ellipses (solid curves) for a selection of placentas from our cohort.}
\label{fig5}
\end{figure}

\subsection{Placental Shape}
\label{sec3.5}

The moment expansion method described in \ref{appendixB} has been used to calculate the best-fit ellipse for the chorionic plate of each placenta in the cohort.  This includes the semi-major and semi-minor axes and the orientation angle. Figure~\ref{fig5} shows a selection of placentas together with their best-fit ellipses.  Most apparent from this figure is that some outlines fit their ellipse quite well.  These correspond to chorionic plates with regular shapes.  For chorionic plates with irregular shapes that have pronounced lobes and other protrusions, the ellipse does not provide as good a fit.  As expected from the discussion in \ref{appendixB}, such placentas have appreciable higher-order moments to account for their irregularities.  In such cases, quantities derived from higher-order moments, such as skewness (third-order) and kurtosis (fourth-order) provide significant additional information about placental shape.  The extent to which the ellipse accounts for the shape of the chorionic plate can also be used to estimate the roughness.  Any region of the chorionic outline that lies within the ellipse or crosses its boundary contributes to the roughness, as is apparent from the definition in (\ref{eqA6}).

\begin{figure}[t!]
\centering
\includegraphics[width=0.9\textwidth]{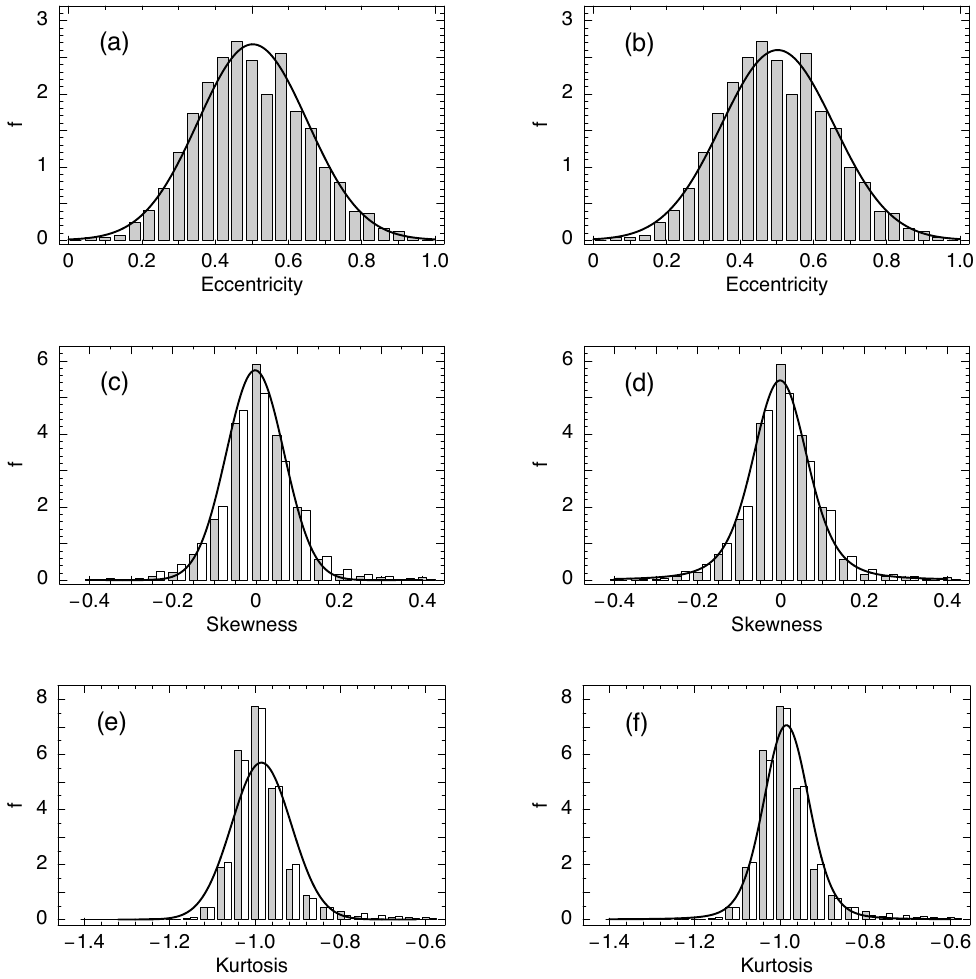}
\caption{Histograms of the eccentricities of the best-fit ellipses compared with (a) a normal and (b) an optimized L\'evy distribution.  Histograms of the skewness of the chorionic plate, projected onto the $x$- and $y$-axes (shown as shaded and unshaded bins, respectively) compared with the (c) normal and (d) an optimized L\'evy distribution.  Histograms of the kurtosis of the chorionic plate, projected onto the $x$- and $y$-axes (shown as shaded and unshaded bins, respectively) compared with the (e) normal and (f) an optimized L\'evy distribution.  The fits in (c)--(f) were obtained by averaging over the $x$- and $y$-projections of each quantity.}
\label{fig6}
\end{figure}

Figure~\ref{fig6}(a) shows the histogram of the eccentricities with a superimposed normal distribution that has the same mean and variance.  The bin sizes were 0.04 for the eccentricity and kurtosis and 0.05 for the skewness.  Figure~\ref{fig6}(a,c,e) shows comparisons of these histograms with normal distributions that have the same mean and standard deviation as the data, while Fig.~\ref{fig6}(b,d,f) shows comparisons with optimized L\'evy distributions using the procedure described in Sec.~\ref{sec3.2}.  The optimized parameters are $\alpha=2$ and $\gamma=0.012$ for the eccentricity, $\alpha=1.6$ and $\gamma=0.009$ for the skewness, and $\alpha=1.75$ and $\gamma=0.0036$ for the kurtosis, with the same error bars as in Sec.~\ref{sec3.2}.  For the skewness and kurtosis, the fits were carried out for the averages over the $x$- and $y$-projections of each quantity.

\section{Discussion}
\label{sec4}

There are large variations in the characteristics of  mature placental shapes. How do these variations arise? The uterine environment plays a part, for example, in cases where ``trophotropism'' suggests that the placenta can differentially grow, effectively migrating to a more suitable location in the uterus.  However, there may also be manifestations of randomness within placental growth. The chorionic plate outline of the mature placenta reflects, in essence, a summary of the effects of all factors that can impact the lateral expansion of the chorionic surface.  Identifying whether a placental measure follows one of the distributions in \ref{appendixC} or another distribution is important for assessing the statistical properties of lateral growth or the processes underlying growth. The results we described in the preceding section are summarized in Table~\ref{table2}.

\begin{table}[t!]
\caption{Summary of best-fit distributions for the measures in Table~\ref{table1}, as well as the distance between the centroid and the umbilical cord insertion.}
\label{table2}
\bigskip
\centering\small
\newcommand\T{\rule{0pt}{2.6ex}}
\newcommand\B{\rule[-1.2ex]{0pt}{0pt}}
\newcommand\TT{\rule{0pt}{4ex}}
\newcommand\BB{\rule[-3ex]{0pt}{0pt}}
\begin{tabular}{|c|c|c|}
\hline
Measure & Best Fit & Comments \T\B\\ 
\hline\hline
\multirow{3}{*}{Area} & \multirow{3}{*}{L\'evy} & Normal distribution for small \TT\\
& &  values, L\'evy distribution for \\
& &  moderate to large values \BB\\
\hline
Roughness & lognormal & power law (``heavy'') tail \TT\BB\\
\hline
Correlation & \multirow{2}{*}{lognormal} & \multirow{2}{*}{power law (``heavy'') tail} \TT\\
Function & & \BB\\
\hline
Centroid-- & \multirow{3}{*}{lognormal} & \multirow{3}{*}{poor fit near peak of histogram} \TT\\
Umbilical Cord & & \\
Distance & & \BB\\
\hline
\multirow{2}{*}{Eccentricity} & \multirow{2}{*}{normal} & optimized L\'evy distribution \TT\\
& & also yields normal distribution \BB\\
\hline
Skewness & L\'evy & average over $x$- and $y$- directions \TT\BB\\
\hline
Kurtosis & L\'evy & average over $x$- and $y$- directions \TT\BB\\
\hline
\end{tabular}
\end{table}

\subsection{Chorionic Plate Area}
\label{sec4.1}

The Gaussian and lognormal distributions provide good accounts of the gross shape of the histogram of chorionic plate areas, but underestimate the peak near the mean [Fig.~\ref{fig2}(a)]. The larger positive deviations from the mean of the area are better described by the lognormal distribution, but the L\'evy distribution provides a discernibly better overall fit to the entire histogram than either the normal or lognormal distributions [Fig.~\ref{fig2}(c)]. In particular, the L\'evy distribution gives a much better account of the peak near the mean and at large positive deviations from the mean, where the decay is much slower than for the normal distribution. This can be seen directly in Fig.~\ref{fig2}(b,d), where we plot the logarithm of the frequencies in Figs.~\ref{fig2}(a,c) against the normalized area. In this coordinate system, the normal distribution appears as an inverted parabola, as follows from (\ref{eqC1}). Figure~\ref{fig2}(b,d) clearly shows that the lognormal distribution provides a good fit for moderate positive deviations from the mean, but that the L\'evy distribution provides a much better fit at all large positive deviations from the mean. The normal distribution, however, provides a better description of the data at values below the mean.

To appreciate the consequences of the better fit provided by the L\'evy distribution, we return to the notion of a random walk \cite{feller68}, where a ``walker'' takes small sequential steps to the left or right, each chosen randomly with equal probability. As previously noted, as the number of steps increases, the distribution of possible distances from the walkerÕs initial position approaches a normal distribution. L\'evy distributions arise from random walks with step sizes chosen from a distribution for which step sizes decay as a power law for large step lengths.  Hence, the likelihood of a large step is much greater than for a random walk. This has the effect of enhancing the rate of displacement displacement compared to a random walk. The fits in Fig.~\ref{fig2} thereby suggest that placentas whose chorionic plate area is much smaller than the mean, which follow a normal distribution, developed by a series of small independent random steps. Placentas with a chorionic area that is much larger than the mean, however, developed by large steps, or a series of smaller correlated steps. In either case, the growth of placentas with a large chorionic area is manifestly inconsistent with normal behavior.

\subsection{Perimeter Roughness and Correlation Function}
\label{sec4.2}

The skewed histograms in Fig.~\ref{fig3} mean that symmetric distributions are not appropriate, so we have focused on the lognormal distribution. This distribution provides a reasonable fit to the data, though there are evident discrepancies, especially for small values of the correlation function. However, as the semi-logarithmic plots in Fig.~\ref{fig3}(b,d,f) show, while the lognormal distribution accurately accounts for moderate positive deviations from the mean, extreme deviations (the ÒtailsÓ) show systematic differences from this distribution.

\begin{figure}[t!]
\centering
\includegraphics[width=0.9\textwidth]{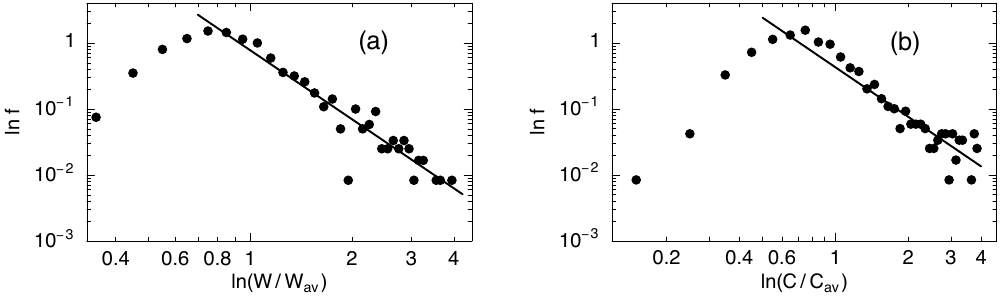}
\caption{Log-log plots of the histograms in Fig.~\ref{fig3}(a,c), with the bin frequencies represented as dots. The lines in each panel represent an optimized linear fit to the tails of the distributions, with slopes of $-3.47$ and $-2.49$, respectively. The quality of these fits suggest that the tails of the corresponding distributions have a power law decay, as in (\ref{eqC6}).}
\label{fig7}
\end{figure}

An analysis of this regime is carried out in Fig.~\ref{fig8}, where we show log-log plots of the data in Figs.~\ref{fig2}(a,c,e) together with a linear fit to the tails of each distribution.  Bearing in mind that there are fewer placentas for extreme values, so the scatter in the data is correspondingly greater than for smaller values, the linear fits provide an acceptable account of these tails. The significance of this becomes apparent when we refer to the discussion in \ref{appendixC} and, in particular, the power law behavior of the tails of the L\'evy distribution in (\ref{eqC6}). The linear fits in Fig.~\ref{fig7} show that the tails of these histograms are indeed consistent with a power law decay. Although this is indicative of the wild fluctuations associated with L\'evy distributions, we have not been able to fit a L\'evy distribution to the entire range of the data. Nevertheless, the analysis in Fig.~\ref{fig7} is very suggestive.  But we conclude this discussion with a word of caution. Linear fits to log-log plots typically rely on several decades (i.e.~powers of ten on a log-log plot) of data to enable a firm conclusion to be drawn about the existence of power law tails. Our analysis is based on less than half a decade, which is the nature of the measures we are using, so our conclusions must be tempered accordingly.

\subsection{Distance between the Centroid and the Umbilical Cord Insertion}
\label{sec4.3}

The lognormal distribution in Fig.~\ref{fig4} provides a good account of the profile of the histogram of distances between the centroid and the umbilical cord insertion -- only the main peak of the histogram is overestimated by approximately 10\%.   Even more significant is the fit in Fig.~\ref{fig6}(b), which shows that the lognormal distribution provides an excellent account of the tail of the histogram.  Hence, we conclude that the distribution associated with this quantity, when measured in mature placentas across a cohort, is the cumulative result of small multiplicative random steps.  This, in turn, leads to two further considerations.  Consider first the fact that a vasaculogenic zone is already evident at the 5th week of development \cite{kliman98}.  Thus, in the early stages of development, we expect that the centroid of the developing chorionic plate and umbilical cord insertion are strongly correlated.  However, as further development occurs, random factors diminish this correlation, eventually producing the uncorrelated behavior seen in Fig.~\ref{fig4}.  Why is the distribution lognormal, rather than normal?  Chemical reactions and, by extension, biochemical reactions, are inherently multiplicative processes \cite{limpert01} because concentrations of particular species must be simultaneously present at a specific location for development to occur.  The amount of each quantity, which varies across the placenta, determines the extent to which development occurs.  The comparisons in Fig.~\ref{fig4} suggests that these spatial variations are random.

\subsection{Placental Shape}
\label{sec4.4}

The most striking result in Fig.~\ref{fig6} is how well the normal distribution accounts for the eccentricities of the best-fit ellipses across the cohort.  This is confirmed by the optimized L\'evy distribution, which has $(\alpha=2)$ and a standard deviation of $\sqrt{2\gamma}=0.0155$, the latter comparing well with the value $\sigma=0.0149$ obtained directly from the data.  Hence, the eccentricity may be regarded as being normally distributed.  The shape of the skewness is also described moderately well by a normal distribution, though the large deviations from the mean are better accounted for by the L\'evy distribution.
For the kurtosis, the normal distribution accounts for the width of the distribution, but underestimates the height of the peak near the mean and, of course, does not account for the occurrence of large deviations from the mean.  Here, the L\'evy distribution provides the superior description.

\section{Summary}
\label{sec5}

Placental weight is a standard measure of placental development and is often used as a primary indicator of fetal health.  But weight is just one way that factors affect the developmental history of a placenta.  Other measures have been presented before \cite{salafia05} and are revisited here in light of their distributions.  Working from interpolations between data points obtained from digitized images of the cohort described in Ref.~\cite{kaufman03}, we have calculated several measures of chorionic plate morphology, including its area, the roughness and correlation function of the outline, the distance between the centroid and the umbilical cord insertion, and several shape parameters.

Our focus here is determining the extent to which the distributions of placental measures are described by normal or lognormal distributions, in other words, the extent to which the fluctuations of these measures result from the sum or product, respectively, of relatively {\it independent} factors. In fact, we found that normal distributions provide an accurate account only of the distribution of the eccentricities of the best-fit ellipses.  Taken together, the results presented here demonstrate how an analysis of a cohort can reveal fundamental aspects in the development of placentas. The deviations from normal or log-normal behavior, in particular, provide the most direct indication of the presence of correlations in the development of the placenta. Large deviations from mean behavior are not simply the result of mild independent fluctuations, as normal or lognormal distributions would imply, but embody the wild fluctuations that lead to power law decay. While we have focused entirely on geometric and morphological features in this paper, other characteristics of the chorionic plate would also benefit from our analysis, especially those which take account of vasculature. Such studies are in progress and will be reported in a future publication.

\newpage

\appendix 

\section{Fourier Series for the Chorionic Plate Outline}
\label{appendixA}

The chorionic plate outline is represented by a set of points with coordinates $(x_k,y_k)$, for $k=1,\ldots,N$ obtained from the digitized images (Sec.~\ref{sec2.1}). To eliminate any bias in the data, we first calculate the coordinates $(x_c,y_c)$ of the centroid by taking the average of each perimeter coordinate:
\begin{equation}
x_c={1\over N}\sum_{k=1}^N x_k\, ,\qquad
y_c={1\over N}\sum_{k=1}^N y_k\, .
\end{equation}
The centroid is taken as the origin of coordinates for the points along the chorionic outline.  The radius $r$ is specified in terms of the length $s$ along the perimeter, which has length $L$.  The Fourier series for $r(s)$ is
\begin{equation}
r(s)=r_{\rm av}+\sum_{n=1}^N\biggl[a_n\cos\biggl({2\pi sn\over L}\biggr)+
b_n\sin\biggl({2\pi sn\over L}\biggr)\biggr]\, ,
\label{eqA3}
\end{equation}
where the Fourier coefficients are
\begin{align}
\label{eqA4}
a_n&={2\over L}\sum_{k=1}^N r_k\cos\biggl({2\pi s_k n\over L}\biggr)\, ,\\
\noalign{\vskip3pt}
b_n&={2\over L}\sum_{k=1}^N r_k\sin\biggl({2\pi s_k n\over L}\biggr)\, ,
\label{eqA5}
\end{align}
in which $r_k$ is the radius of the $k$th data point at a distance $s_k$ along the perimeter. The corresponding series for the coordinates $(x(s),y(s))$ of the perimeter are of the same form as (\ref{eqA3}), but with $x_k$ and $y_k$ in turn replacing $r_k$ in (\ref{eqA4}) and (\ref{eqA5}).

The interpolation of the chorionic plate outline can be used to calculate several measures associated with the deviations of this outline from circularity.  The roughness $W$ of this outline is defined as the
root-mean-squared deviations from its average radius $r_{\rm av}$:
\begin{equation}
W=\biggl\{{1\over L}\int_0^L\bigl[r(s)-r_{\rm av}\bigr]^2\,ds\biggr\}^{1/2}\, ,
\label{eqA6}
\end{equation}
in which $r(s)$ is the distance from the centroid to the chorionic plate outline at a point $s$ along the outline and $L$ is the length of the outline.  The roughness is expressed in terms of the coefficients in (\ref{eqA4}) and (\ref{eqA5}) as
\begin{equation}
W=\biggl[{1\over2}\sum_{n=1}^\infty\bigl(a_n^2+b_n^2\bigr)\biggr]^{1/2}\, .
\label{eqA7}
\end{equation}

The correlation function $C(s)$, defined as
\begin{equation}
C(s)=\biggl\{{1\over L}\int_0^L\bigl[r(s+t)-r(t)\bigr]^2\,dt\biggr\}^{1/2}\, ,
\label{eqA8}
\end{equation}
is the standard deviation of pairs of radii on the chorionic plate outline as a function of their separation, is expressed in terms of the coefficients in (\ref{eqA4}) and (\ref{eqA5}) as
\begin{equation}
C(s)=\biggl[2\sum_{n=1}^\infty\bigl(a_n^2+b_n^2\bigr)\sin^2\biggl({\pi sn\over L}\biggr)\biggr]^{1/2}\, .
\label{eqA9}
\end{equation}
The relation between the correlation function and the roughness can be obtained directly from (\ref{eqA7}) and (\ref{eqA9}):
\begin{equation}
\int_0^LC^2(s)\,ds=L\sum_{n=1}^N\bigl(a_n^2+b_n^2\bigr)=2LW^2\, ,
\label{eqA10}
\end{equation}
so the correlation function corresponds to a roughness that is spatially resolved along the chorionic plate outline.

\section{Moments of Chorionic Plate Shape}
\label{appendixB}

An alternative to the contour-based analysis of chorionic plate shape using the Fourier series in \ref{appendixA} is the area-based approach of moments.  We define a function $f(x,y)$ that takes the value 1 within the chorionic plate area and the value 0 outside this area.  The $(p,q)$th moment $\mu_{p,q}$ of the enclosed area is defined as
\begin{equation}
\mu_{p.q}=\iint x^p y^q f(x,y)\,dx\,dy\, ,
\end{equation}
where $p,q=0,1,2,\cdots$.  If all of the moments are calculated, then the original shape can be restored.  In practice, only lower-order moments, for which $p+q\le4$ are typically used.

The zeroth-order moment $\mu_{0,0}$ determines the area $A$ of the chorionic plate,
\begin{equation}
A=\mu_{0,0}=\iint f(x,y)\,dx\,dy\, ,
\end{equation}
and the coordinates $(x_c.y_c)$ of the centroid are expressed in terms of $\mu_{0,0}$ the first-order moments $\mu_{1,0}$ and $\mu_{0,1}$ as
\begin{align}
x_c&={1\over A}\iint x\,f(x,y)\,dx\,dy={\mu_{1,0}\over\mu_{0,0}}\, ,\\
\noalign{\vskip3pt}
y_c&={1\over A}\iint y\,f(x,y)\,dx\,dy={\mu_{0,1}\over\mu_{0,0}}\, .
\end{align}
The zeroth- and second-order moments determine the best-fit ellipse.  

This ellipse is centered at the centroid of the chorionic plate and its semi-major and semi-minor axes $a$ and $b$, respectively, are the perpendicular lines that pass through the centroid for which the second-order central moments about these lines are maximum and minimum, respectively.  The semi-major and semi-minor axes are given by \cite{teague80}
\begin{align}
a=\sqrt{2}\biggl\{{\mu_{2,0}+\mu_{0,2}+\bigl[(\mu_{2,0}-\mu_{0,2})^2+4\mu_{1,1}^2\bigr]^{1/2}\over\mu_{0,0}}\biggr\}^{1/2}\, ,\\
\noalign{\vskip3pt}
b=\sqrt{2}\biggl\{{\mu_{2,0}+\mu_{0,2}-\bigl[(\mu_{2,0}-\mu_{0,2})^2+4\mu_{1,1}^2\bigr]^{1/2}\over\mu_{0,0}}\biggr\}^{1/2}\, ,
\end{align}
where the tilt angle $\phi$ of the ellipse, measured counterclockwise with respect to the original coordinate axes, is \cite{teague80}
\begin{equation}
\phi={1\over2}\tan^{-1}\biggl({2\mu_{1,1}\over\mu_{2,0}-\mu_{0,2}}\biggr)\, .
\end{equation}
The convention is that $\phi$ is the angle between the $x$-axis and the semi-major axis, where, by definition, $a\ge b$.  The eccentricity $e$ of the best-fit ellipse is given by the usual formula:
\begin{equation}
e=\sqrt{1-{b^2\over a^2}}\, .
\end{equation}

Higher-order moments include quantities such as skewness and kurtosis of $x$ and $y$ projections of the placental shape (for example, the $x$-projection is the image obtained by summing over all pixels in the $x$-direction). Expressions for these quantities are compiled in Table~\ref{table1}.

\section{Probability Density Functions}
\label{appendixC}

The probability density function $p(x)$ of a continuous random variable represents the relative likelihood that the random variable occurs at a given point $x$. The probability density function is nonnegative, and its integral over all possible values of $x$ is equal to one.  The probability density of the normal distribution is
\begin{equation}
p_1(x;\mu,\sigma)={1\over\sqrt{2\pi\sigma^2}}\exp\biggl[-{(x-\mu)^2\over2\sigma^2}\biggr]\, ,
\label{eqC1}
\end{equation}
in which $\mu$ is the mean $\sigma$ the standard deviation.  The corresponding quantity for the lognormal distribution is
\begin{equation}
p_2(x;\mu,\sigma)={1\over x\sqrt{2\pi\sigma^2}}\exp\biggl[-{(\ln x-\mu)^2\over2\sigma^2}\biggr]\, .
\end{equation}
where $\mu$ is the mean and $\sigma$ the standard deviation for $\ln x$.  These are related to the mean $\mu^\prime$ and variance $\sigma^{\prime\,2}$ of a random variable that is log-normally distributed by
\begin{align}
\label{eqC3}
\mu&=\ln(\mu^\prime)-{\textstyle{1\over2}}\ln\biggl(1+{\sigma^{\prime\,2}\over\mu^{\prime\,2}}\biggr)\, ,\\
\noalign{\vskip3pt}
\sigma^2&=\ln\biggl(1+{\sigma^{\prime\,2}\over\mu^{\prime\,2}}\biggr)\, .
\label{eqC4}
\end{align}
The probability density function of the L\'evy distribution
\begin{equation}
p_3(x;\alpha,\gamma)={1\over\pi}\int_0^\infty e^{-\gamma k^\alpha}\cos(kx)\,dk\, ,
\label{eqC5}
\end{equation}
where $0<\alpha\le2$ and $\gamma>0$ is a width parameter. The L\'evy distribution is known in closed form only for $\alpha=1$ and $\alpha=2$, with the latter yielding the normal distribution in the form
\begin{equation}
p_3(x;2,\gamma)={1\over\sqrt{4\pi\gamma}}\exp\biggl(-{x^2\over4\gamma}\biggr)\, ,
\label{eqC6}
\end{equation}
which is a normal distribution with $\mu=0$ and $\sigma^2=2\gamma$.  In all other cases the L\'evy distribution must be evaluated numerically.

One of the most important characteristics of L\'evy distributions is that the probability density of extreme variations of a random variable follows a power law:
\begin{equation}
p_3(x;\alpha,\gamma)\to |x|^{-\alpha-1}\quad\mbox{as}\quad |x|\to\infty\, .
\label{eqC7}
\end{equation}

\newpage

\bibliographystyle{elsarticle-num}


\end{document}